# New Localizable Entanglement

Abbaas Sabour [1*,], Fereydoon Khazali [2,], Soghra Ghanavati [1]

[1] Department of Physics, Om. C., Islamic Azad University, Omidiyeh, Iran.
[2] Department of Chemistry, Om. C., Islamic Azad University, Omidiyeh, Iran.

**Abstract**

In this study, we have addressed an ambiguity in the concept of localizable entanglement (*LE*) introduced by Verstraete et al in 2004. By doing so, we have proposed and explored a unique form of this entanglement, called new localizable entanglement (*NLE*). We have shown that *NLE* is always less than or equal to *LE*. Additionally, we have demonstrated that for systems with three components, *NLE* does not differ significantly from *LE*. However, when the number of components increases to four, there is a possibility of significant differences between the two methods. Furthermore, as the number of components increases further, this difference becomes slightly more pronounced. It appears that the classical correlation, which is the lower bound for *LE*, is also a lower bound for *NLE*.

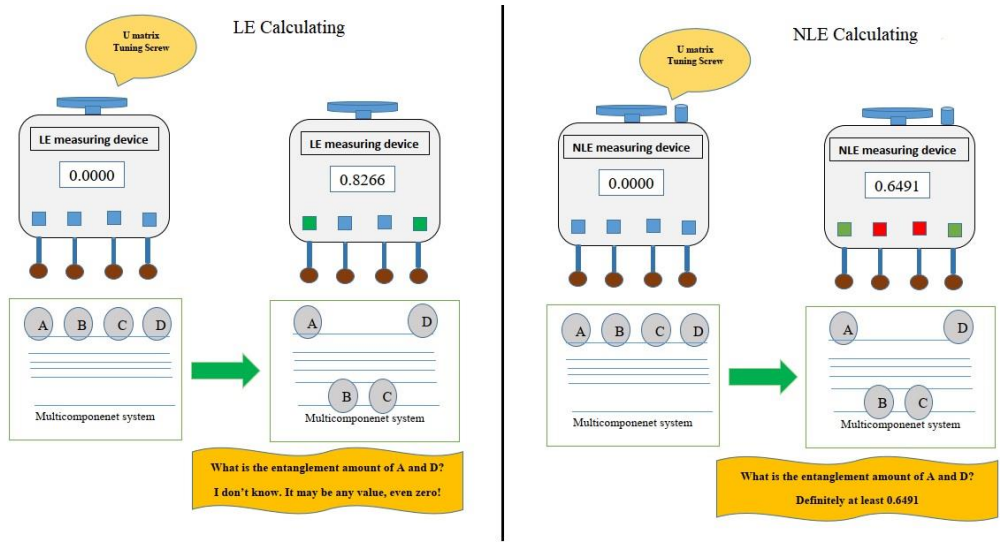

## Introduction

In many fields of quantum information, the creation and distribution of entangled states are very important[1,2], because it is a key element for performing some tasks of quantum information, such as teleportation or quantum computing [3]. In this sense, many-particle quantum states that appear naturally in many physical systems can be considered entanglement resources [4]. Hence, it is desirable to find ways to describe and quantify entanglement in multipartite systems. In many cases, the goal is to calculate the entanglement between two components of a multicomponent system [5]. But the entanglement of multicomponent systems has always been of interest, and a number of measures have been proposed for this issue, some of which are based on calculating the entanglement between two components for all possible pairs in the multi-component system [6]. The basic question raised here is that if all the components of a multi-component system are entangled, how can be calculated the entanglement between only two components in a multicomponent system? To answer this question, it is necessary to describe the state of the subsystems. The oldest and most common method to obtain the state of a subsystem, which was first used by Dirac in 1930, is the use of the reduced density operator[7]. Using this method, in general, for a two-component subsystem, a mixed state is obtained even when the state of the whole system is pure (in continuing, a simple example is given.). It is clear that the density operator of a mixed state cannot be thought of as consisting of a certain ensemble of pure states with a certain probability since there are countless ensembles that create a particular density operator. Of course, this method gives an unambiguous answer for product states. In another method to calculate the entanglement between two components of a multicomponent system, instead of the partial tracing method, the performing measurement on the rest of the multi-component system is used. In the latter method, there is no need to assume the existence of a specific density operator independent of external factors and to calculate the entanglement of the two-component subsystem by performing measurements on the rest of the multicomponent system, the wave function is collapsed into one of its eigenstates, so the entanglement between the subsystem and the rest of the system is removed [8-10]. Of course, by adjusting the measuring devices, the maximum average entanglement between the two components of the system can be obtained. The adjustment of the measuring devices is done by applying a suitable unitary transformation to the measuring operators, which is equivalent to applying the adjoint of this transformation to the system state. If the measurements are limited to only local measurements, localizable entanglement (*LE*), will be the result [11]. Local measurements are divided into three categories according to the degree of universality of the measuring facilities, including local operations and classical communication (LOCC), positive operator-valued measure (POVM), and projective von-Neumann measurements (PM) [12]. The unitary transformation matrix representation for a qubit can be shown as follows[13]

$$u_{\theta_1,\theta_2,\alpha,\beta} := \begin{bmatrix} e^{i\theta_1} & 0 \\ 0 & e^{i\theta_2} \end{bmatrix} \begin{bmatrix} \cos\alpha & -\sin\alpha e^{-i\beta} \\ \sin\alpha e^{i\beta} & \cos\alpha \end{bmatrix} \qquad (1)$$

---

* Corresponding author e-mail address: ab.sabour@iau.ac.ir

The present paper begins with a comprehensive review of the commonly used *LE* method for pure states, including a detailed explanation of its lower bound. Next, a novel approach referred to as the new localizable entanglement (*NLE*) is introduced and compared with the conventional method. To provide further clarity, illustrative examples are presented. In this paper, for the sake of simplicity, projective von-Neumann measurements (PM) are used, as well as, special attention will be paid to four-qubit systems because it shows a bigger gap between the values of *LE* and *NLE* compared to smaller systems.

**Localizable Entanglement**

In reference[11,14] the *LE* which calculates the entanglement between two components of a multipartite system is defined. The *LE* is the maximum entanglement that can be localized, on average, between two parties of a multipartite system, by performing local measurements on the other parties. Therefore, with this definition, *LE* has an obvious physical concept and is an operational approach. The reason for the selection of averaging (rather than selecting a maximally entangled state) in this definition is that each local measurement transforms the system into a specific state with a certain probability, and it is possible for a state with high entanglement to occur with a very small probability. Note that for calculating *LE*, tracing is not performed on the components, but rather measurements are made. This means that if the overall state of the system is a pure state, and the measurements made on the rest of the system components are of PM measurement types, the subsystem state collapses deterministically and completely to a specific pure and determinate state. To illustrate the fundamental difference between tracing and measuring, a comparison between the results of these two methods is given for the GHZ state as a simple example. the GHZ state is defined as follows [15]

$$|GHZ\rangle = \frac{1}{\sqrt{2}}\left(|0^A 0^B 0^C\rangle + |1^A 1^B 1^C\rangle\right) \tag{2}$$

For the GHZ state which has the maximum total entanglement for the 3-qubit systems, it can easily be demonstrated that the reduced density operator of both qubits in the GHZ system, obtained by partial tracing over the third qubit, exhibits no entanglement at all, whereas properly setting up the measurement device for local measurements on each qubit of the GHZ system, the entanglement between the other two qubits is maximized.

$$If: PM_{\pm}^C = \frac{1}{\sqrt{2}}\left(|0^C\rangle \pm |1^C\rangle\right)\frac{1}{\sqrt{2}}\left(\langle 0^C| \pm \langle 1^C|\right) := D_{\pm}^C \tag{3}$$

$$PM_{\pm} = 1^A \otimes 1^B \otimes D_{\pm}^C \Rightarrow PM_{\pm}\left(\rho^{ABC} = |GHZ\rangle\langle GHZ|\right)PM_{\pm}^{\dagger} = \left(\rho^{AB} = |\varphi_{\pm}^{AB}\rangle\langle \varphi_{\pm}^{AB}|\right) \otimes D_{\pm}^C \tag{4}$$

That here $|\varphi_{\pm}^{AB}\rangle = \frac{1}{\sqrt{2}}\left(|0^A 0^B\rangle \pm |1^A 1^B\rangle\right)$ represents parallel Bell states with the entanglement of one[16]. An exact value of localizable entanglement for a general three-qubit was obtained by Gao et. al[17].

The method of *LE* presents a clear physical interpretation of the degree of entanglement between two components. Furthermore, it establishes a close connection between the entanglement and classical correlation functions ($Q^{AB}$) of the two components *A* and *B* and has demonstrated that $Q^{AB}$ consistently yield lower bounds to the *LE*[18]. Where in:

$$If: Q_{\hat{\alpha},\hat{\beta}}^{AB}\left(|\psi\rangle\right) := \langle\psi|\sigma_{\hat{\alpha}}^A \otimes \sigma_{\hat{\beta}}^B|\psi\rangle - \langle\psi|\sigma_{\hat{\alpha}}^A|\psi\rangle\langle\psi|\sigma_{\hat{\beta}}^B|\psi\rangle \, and \, \sigma_{\hat{\alpha}} := \left(\hat{i}\sigma_x + \hat{j}\sigma_y + \hat{k}\sigma_z\right).\hat{\alpha}$$
$$\Rightarrow Q^{AB}\left(|\psi\rangle\right) := \underset{\hat{\alpha},\hat{\beta}}{Max}\left(Q_{\hat{\alpha},\hat{\beta}}^{AB}\left(|\psi\rangle\right)\right) \tag{5}$$

The mathematical procedure for determining the *LE* between two specified sites of a 4-qubit system, such as between sites 1 and 4, can be expressed as follows: Assuming the $|\psi\rangle$ represents the state of the four-spin system, the adjustment of the measuring devices by rotation for the purpose of conducting a measurement on the remaining two sites (2 and 3) can be represented as a local unitary transformation $U_{2,3} = 1 \otimes u_2 \otimes u_3 \otimes 1$ applied to the system state. Measuring the results of $j_2$ and $j_3$ on sites 2 and 3 is equivalent to applying the projective operator $N^{j_2 j_3} = 1 \otimes |j_2\rangle\langle j_2| \otimes |j_3\rangle\langle j_3| \otimes 1$.

Consequently, by locally adjusting and performing measurements on the two sites 2 and 3, they can collapse into specific eigenstates corresponding to particular eigenvalues, while only sites 1 and 4 remain entangled, and the system goes to the state (un-normalized) $|\tilde{\psi}_{j_2 j_3}^{U_{2,3}}\rangle = N^{j_2 j_3} U_{2,3} |\psi\rangle$ with a probability of $P_{j_2 j_3}^{U_{2,3}} = \langle \tilde{\psi}_{j_2 j_3}^{U_{2,3}} | \tilde{\psi}_{j_2 j_3}^{U_{2,3}} \rangle$, therefore, the $LE_{1,4}$ can be calculated using the following equation.

$$LE_{1,4} = \underset{U_{2,3}}{Max} \sum_{j_2, j_3 = 0}^{1} P_{j_2 j_3}^{U_{2,3}} C\left(\left|\psi_{j_2 j_3}^{U_{2,3}}\right\rangle\right) \tag{6}$$

In equation (6), $C$ is the commonly used measure of concurrence for calculating entanglement[19,20], which is calculated as follows for pure states: $C(a|11\rangle + b|10\rangle + c|01\rangle + d|00\rangle) = 2|ad - bc|$

The operational concept of this common measure was first introduced by one of the authors of this paper[21]. Maximization is performed over all possible settings of the measurement devices (mathematically, it is maximization over all parameters that determine the unitary operators applied to sites 2 and 3). Clearly, this approach has a significant advantage over the partial trace method in terms of having a physical interpretation, as there exists an operational procedure to calculate it. However, from a mathematical perspective, due to the necessary maximizations and complex definitions, its analytic calculation is very challenging.

**New Localizable Entanglement**

If the components other than the two components desired to calculate the entanglement of a multi-component system, which is in a certain pure state, are subjected to the local PM measurement, practically, finally, the state of the two desired components collapses into a specific non-local pure state (not an ensemble of different pure states), which depends on the setting of the measuring devices and their outputs. In the definition of *LE*, the maximum average entanglement of such states is calculated after proper adjustment of the measuring devices, which in practice must be obtained after a large number of measurements on similar systems, because in just doing one measurement, the system collapses into only one pure state, which, its entanglement apart from the possibility of their occurrence can have any value from 0 to 1. Although this seems obvious, to demonstrate it in practice, we considered a general three-qubit system and created a large number of pure normal states with random components, and then we grouped them in such a way that the states of each category produce a specific value of *LE* (within a small interval of about 0.01) for two specific components of the system. After checking the minimum and maximum entanglement of the collapsed states in each category, the result was as follows: The higher the number of initial random states (we applied once around 20,000 states and another time around 200,000 states), the closer these minimums and maximums got to zero and one, respectively, regardless of the value of *LE*.

An important question arises: Is it useful to know the *LE* value between two components when the entanglement value obtained after each measurement could be any value? As mentioned earlier, the method of reducing the density operator by taking a partial trace converts the system state into a mixed state, and the density matrix of a mixed state cannot be thought of as consisting of a specific ensemble of pure states with a certain probability of occurrence. However, in reality, a localized measurement with a certain probability collapses the subsystem to a specific state (and if the system is pure, to a pure state). By examining the result of the experiment, we can infer the state of the subsystem (even if its probability is low). Therefore, a new type of localizable entanglement is introduced based on the previous method of localizable entanglement, which does not suffer from the drawbacks of the original method. To calculate the new localizable entanglement, measurements are first performed on other parts of the system, similar to before, using measurements of any type such as LOCC, POVM, or PM (in this study, only PM measurements were used). To perform the measurement, the operator parameters corresponding to the observable being measured (in fact, local unitary transformation parameters) are adjusted in such a way that the entanglement of the measured state for those two parts is maximized for the entangled state that minimizes the entanglement. The maximum value obtained from this process is defined as the *NLE*. It is clear that any other eigenvalue obtained creates greater entanglement than the value of the *NLE*.

**Relation of *NLE* and *LE***

We now prove that $LE \geq NLE$. Without loss of generality, we present this proof for the case of a four-qubit system. The results can easily be generalized to more general cases:

$$LE_{1,4} = \underset{U_{2,3}}{Max} \sum_{j_2,j_3=0}^{1} P_{j_2 j_3}^{U_{2,3}} C\left(\left|\psi_{j_2 j_3}^{U_{2,3}}\right\rangle\right) \geq \left. \sum_{j_2,j_3=0}^{1} P_{j_2 j_3}^{U_{2,3}} C\left(\left|\psi_{j_2 j_3}^{U_{2,3}}\right\rangle\right) \right|_{\forall U_{2,3}} \tag{7}$$

The above inequality holds true for any $U_{2,3}$, including specific transformations such as the $U_{2,3}^s$ where:

$$Min\left\{C\left(\left|\psi_{0,0}^{U_{2,3}^s}\right\rangle\right), C\left(\left|\psi_{0,1}^{U_{2,3}^s}\right\rangle\right), C\left(\left|\psi_{1,0}^{U_{2,3}^s}\right\rangle\right), C\left(\left|\psi_{1,1}^{U_{2,3}^s}\right\rangle\right)\right\} \geq Min\left\{C\left(\left|\psi_{0,0}^{U_{2,3}}\right\rangle\right), C\left(\left|\psi_{0,1}^{U_{2,3}}\right\rangle\right), C\left(\left|\psi_{1,0}^{U_{2,3}}\right\rangle\right), C\left(\left|\psi_{1,1}^{U_{2,3}}\right\rangle\right)\right\}\Big|_{\forall U_{2,3}} \tag{8}$$

that means a transformation that maximizes the minimum value of entanglement among collapsed states. Now if we define:

$$NLE_{1,4} := \underset{U_{2,3}}{Max}\left(Min\left\{C\left(\left|\psi_{0,0}^{U_{2,3}}\right\rangle\right), C\left(\left|\psi_{0,1}^{U_{2,3}}\right\rangle\right), C\left(\left|\psi_{1,0}^{U_{2,3}}\right\rangle\right), C\left(\left|\psi_{1,1}^{U_{2,3}}\right\rangle\right)\right\}\right) \tag{9}$$

and hence:

$$LE_{1,4} \geq \sum_{j_2,j_3=0}^{1} P_{j_2 j_3}^{U_{2,3}^s} C\left(\left|\psi_{j_2 j_3}^{U_{2,3}^s}\right\rangle\right) \geq \sum_{j_2,j_3=0}^{1} P_{j_2 j_3}^{U_{2,3}^s} NLE_{1,4} = NLE_{1,4} \sum_{j_2,j_3=0}^{1} P_{j_2 j_3}^{U_{2,3}^s} = NLE_{1,4} \tag{10}$$

It can be shown (an example is given following) that these two may not necessarily exhibit similar behavior. Even if the graph of one function is not continuous with respect to a variable such as time and therefore not differentiable, there is no requirement for the other function to exhibit a similar behavior. Now we examine the maximum difference between *LE* and *NLE* in each system. For this purpose, we create many random states (about 100,000 states) for three-qubit, four-qubit, and five-qubit systems, and for each state, we calculate both *LE* and *NLE* values for the same two components. Then, for each system, we selected the state that showed the greatest difference between the LE and the *NLE* values. The values of this maximum difference for each system are reported in Table 1. These values provide a suitable lower bound for the exact maximum difference in each system.

Table 1: calculated values of *MR*, *MD*, and *A* for 3-qubit, 4-qubit, and 5-qubit systems

| N | MR | MD | A |
|---|---|---|---|
| $N=3$ | 0.0327 | 0.0177 | 0.5228 |
| $N=4$ | 0.2147 | 0.1775 | 0.6491 |
| $N=5$ | 0.2821 | 0.2092 | 0.6258 |

$$MR := \underset{|\psi\rangle}{Max} \frac{LE(|\psi\rangle) - NLE(|\psi\rangle)}{NLE(|\psi\rangle)}, MD := \underset{|\psi\rangle}{Max}\left(LE(|\psi\rangle) - NLE(|\psi\rangle)\right), A := NLE(|\psi_{Max\Delta LE}\rangle) \quad (11)$$

where in these relations, *MR* and *MD* respectively represent the maximum relative difference and the maximum absolute difference between *LE* and *NLE*, and *A* indicates the value of *NLE* for the state in which the maximum value of *MD* is obtained.

Based on the values presented in the table, it is evident that the difference between *LE* and *NLE* is insignificant for three-qubit systems, but noticeable for four-qubit systems. Moreover, the difference between *LE* and *NLE* slightly increases for five-qubit systems when compared to four-qubit systems. Given that classical correlation functions provide a lower bound for *LE* [9-11], we also investigated this issue for *NLE* using an example without general proof.

As a constructive example, we investigate a four-part chain system with an Ising model Hamiltonian and no magnetic field [10-12].

$$H = J\left(\sigma_x^1 \otimes \sigma_x^2 + \sigma_x^2 \otimes \sigma_x^3 + \sigma_x^3 \otimes \sigma_x^4\right) \quad (12)$$

where, *J* represents the degree of interaction between neighboring sites, and $\sigma_x^a$ (x = 1, 2, 3) refers to the Pauli matrices. We select the state that exhibits the maximum difference between *LE* and *NLE* at time $t=0$ and obtain their difference over a time interval.

$$|\psi_t\rangle = e^{-i\frac{Ht}{\hbar}}|\psi_0\rangle \quad (13)$$

We also calculate the $Q^{1,4}$ for the same interval and plot both results in Figure (1). Based on the graph, it can be inferred that the *Q* serves as a lower bound for *NLE*.

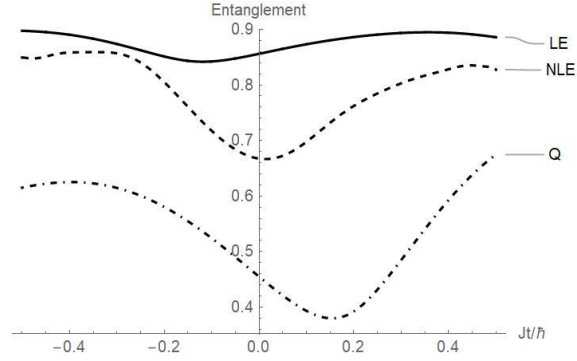

Figure 1: Comparison of *LE*, *NLE*, and *Q* for the 4-qubit system over time ($-0.5\,\hbar/J \geq t \geq 0.5\,\hbar/J$). The time at which the difference between *LE* and *NLE* is maximum has been set to zero. Solid line, dashed line, and dash-dotted line respectively represent *LE*, *NLE*, and *Q*.

**Conclusions and suggestions**

In this study, a different form of localizable entanglement has been introduced, which does not exhibit the ambiguity observed in the definition of *LE*, but shares many similarities with it. The values of *LE* and *NLE* were computed for three-qubit, four-qubit, and five-qubit systems, and their difference was calculated. The results showed that *LE* is always greater than *NLE*. Moreover, it was demonstrated that classical correlation functions provide a lower bound for *NLE*, similar to *LE*. The advantage of *NLE* over *LE* can be expressed as follows. When the *NLE* is calculated between two components of a system, it creates states for those two components that are definitely entangled, at least as much as the calculated *NLE*. However, there is no such connection for the states that are created due to the *LE* calculation, and for a calculated *LE*, the entanglement may even be zero. Considering that, in our view, the physical concept of *NLE* is clearer than that of *LE*, it is recommended that more of its features be explored, particularly in systems with more components or larger dimensions. Additionally, it should be demonstrated that, in general, the classical correlation function provides a lower bound for *NLE*. It is also possible to extend the measurement beyond the von Neumann type and investigate the effect of other local measurements. The results of this study have implications for the creation and distribution of entangled multi-particle states in the field of quantum information.